\RequirePackage{ifpdf}
\documentclass{desyproc}

\begin{document}
%------------------------------------
\title{LHCf: physics results on forward particle production at LHC}

%for single authors the superscripts are optional
\author{{\slshape Oscar Adriani, on behalf of the LHCf Collaboration}\\[1ex]
Universit\`a degli Studi di Firenze and INFN Sezione di Firenze, \\
Via Sansone 1, 50019 Sesto Fiorentino (Fi) , Italy
}

% if the proceedings are available online (e.g. at Indico)
% please enter the contribution ID or file_name below for the DOI
%\contribID{32}
\contribID{adriani}

% TO THE CONFERENCE EDITORS: 
% please update the following information      
% before sending the template to the authors
% \confID{800}  % if the conference is on Indico uncomment this line

\acronym{EDS'13} % if you want the Acronym in the page footer uncomment this line

\maketitle

\begin{abstract}
The LHCf experiment is dedicated to the measurement of very forward particle production in the high energy hadron-hadron collisions at LHC, with the aim of improving the cosmic-ray air shower developments models. The detector has taken data in p-p collisions at $\sqrt s$ = 900 GeV, 2.76 TeV and 7 TeV, and in p/Pb collisions at $\sqrt s$ = 5 TeV. The results of forward production spectra of photons,  neutral pions and neutrons, compared with the models most widely used in the High Energy Cosmic Ray physics, are presented in this paper.
\end{abstract}

\section{The LHCf experiment}

The LHCf experiment is composed by two independent detectors (Arm1 and Arm2), each made by  two sampling and imaging calorimeter towers, composed of 16 tungsten layers, 16 plastic scintillator layers for energy measurement and four position sensitive layers for impact position determination. The position sensitive layers of Arm1 and Arm2 are X-Y scintillating fiber (SciFi) hodoscopes and X-Y silicon strip detectors, respectively. The transverse cross sections of calorimeter towers are 20$\times$20 mm$^2$ and 40$\times$40 mm$^2$ in Arm1 and 25$\times$25 mm$^2$ and 32$\times$32mm$^2$ in Arm2. The detectors are installed 140 m away from the IP1 (Atlas) LHC interaction point. Thanks to the special shape of the beam pipe, the LHCf detectors are able to measure neutral particles in the pseudo-rapidity range from $\eta$ = 8.4 to infinite (zero degree). 
The energy resolution of the detectors for photons and neutrons is 5$\%$ and 35$\%$, respectively, while the position resolution in the neutral particle's reconstruction is better than 200$\mu$m for photons and a few mm for neutrons. More details of the detector performance are reported in~\cite{LHCf_1,LHCf_2}.

\section{Selected results in p-p collisions}

This section summarize some of the results that have been recently published by the LHCf collaboration, concerning $\gamma$ and $\pi^0$ spectra in p-p collisions~\cite{LHCf_3, LHCf_4, LHCf_5}, and shows some preliminary results obtained for the neutron spectra. 
The data used in these analyses have been taken at the beginning of the LHC run, in 2010, in low  luminosity conditions, that are optimal for the LHCf experiment.

\subsection{$\gamma$ spectra}
\begin{figure}[htbp] 
\begin{center}
   \includegraphics[width=1\textwidth]{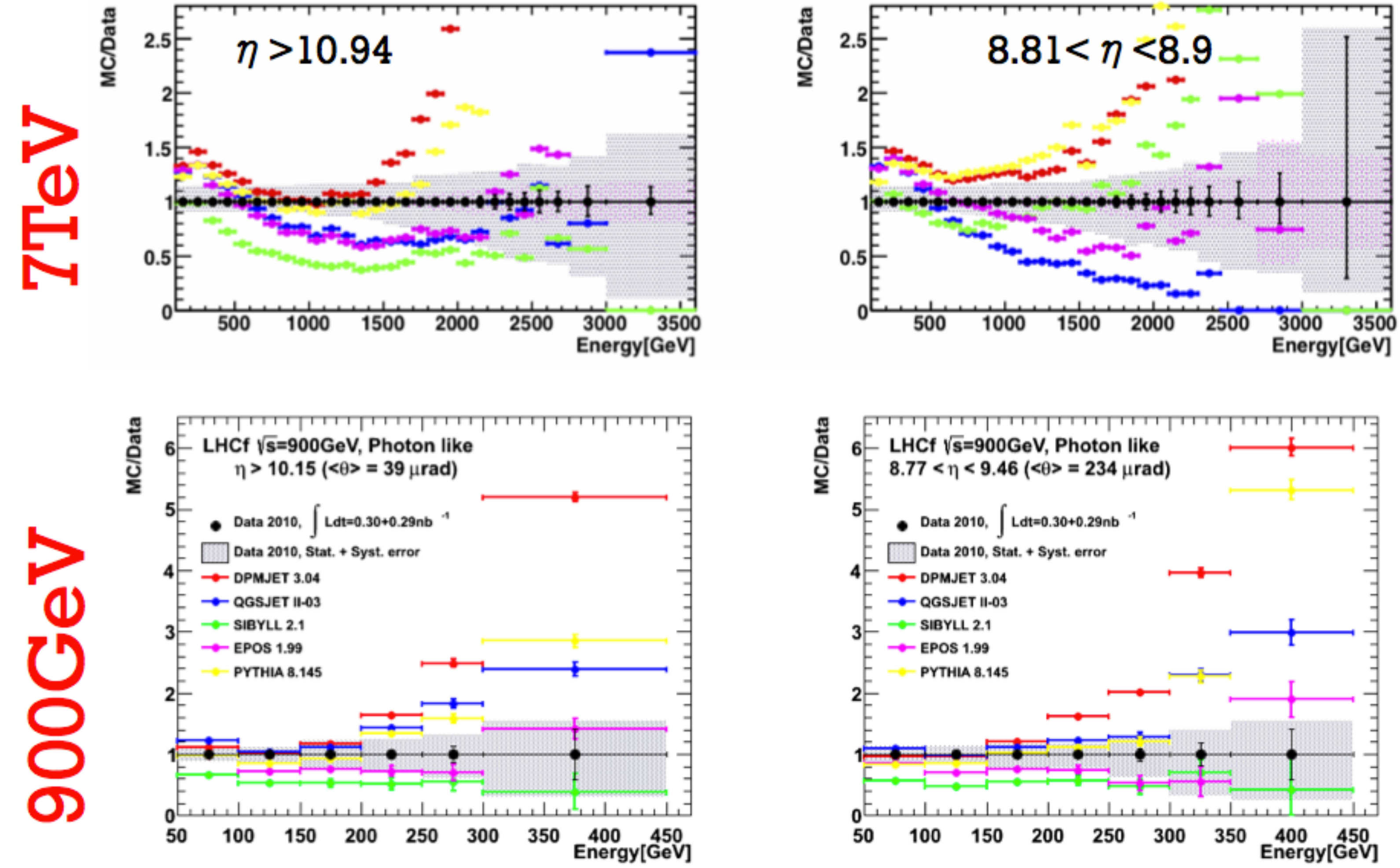}
\caption{Ratio MC/Data for the $\gamma$ spectra obtained by LHCf in the p-p collisions.} 
\label{fig1} 
\end{center}
\end{figure}

Figure~\ref{fig1} shows the ratio between the expectations for the $\gamma$ energy spectra from the models most widely used in the high energy cosmic rays communities (DPMJET3.03, PYTHIA8.145, EPOS1.99, SYBILL2.1, QGSJETII-03) and the $\gamma$ spectra that have been measured in the small ($\eta >10.94$) and large tower ($8.81< \eta < 8.90$) regions. Top figures refer to 7 TeV collisions, while bottom plots show the results for the 900 GeV collisions. 

Although no model is able to reproduce the data perfectly, the LHCf measurements sit on the middle of model predictions. DPMJET3.03 and PYTHIA8.145 show harder spectra than data and EPOS1.99 shows the best overall agreement with data in the tested models. The same tendencies were also found in the $\pi^0$ spectra~\cite{LHCf_5}, as one can naively expect since most of observed photons are produced by decays of neutral pions.

\subsection{Neutron spectra}

The analysis workflow for the neutron energy spectra measurement is similar to the workflow used for the $\gamma$ analysis, described in~\cite{LHCf_3}. The luminosity has been measured with the help of the LHCf front counter rates, properly normalized with the Van der Meer LHC scan. Particle identification has been carried out by looking at the longitudinal shower development, using two dimensional cuts in the L20$_\%$ and L90$_\%$ plane, where L20$_\%$ and L90$_\%$ are the longitudinal depths containing 20$_\%$ and 90$_\%$ of the total deposited energy, respectively. Hit position has been evaluated by using the transverse shower distribution, measured with the position sensitive layers, to optimally correct the energy measured for the leakage effects. 

\begin{figure}[htbp] 
\begin{center}
   \includegraphics[width=1\textwidth]{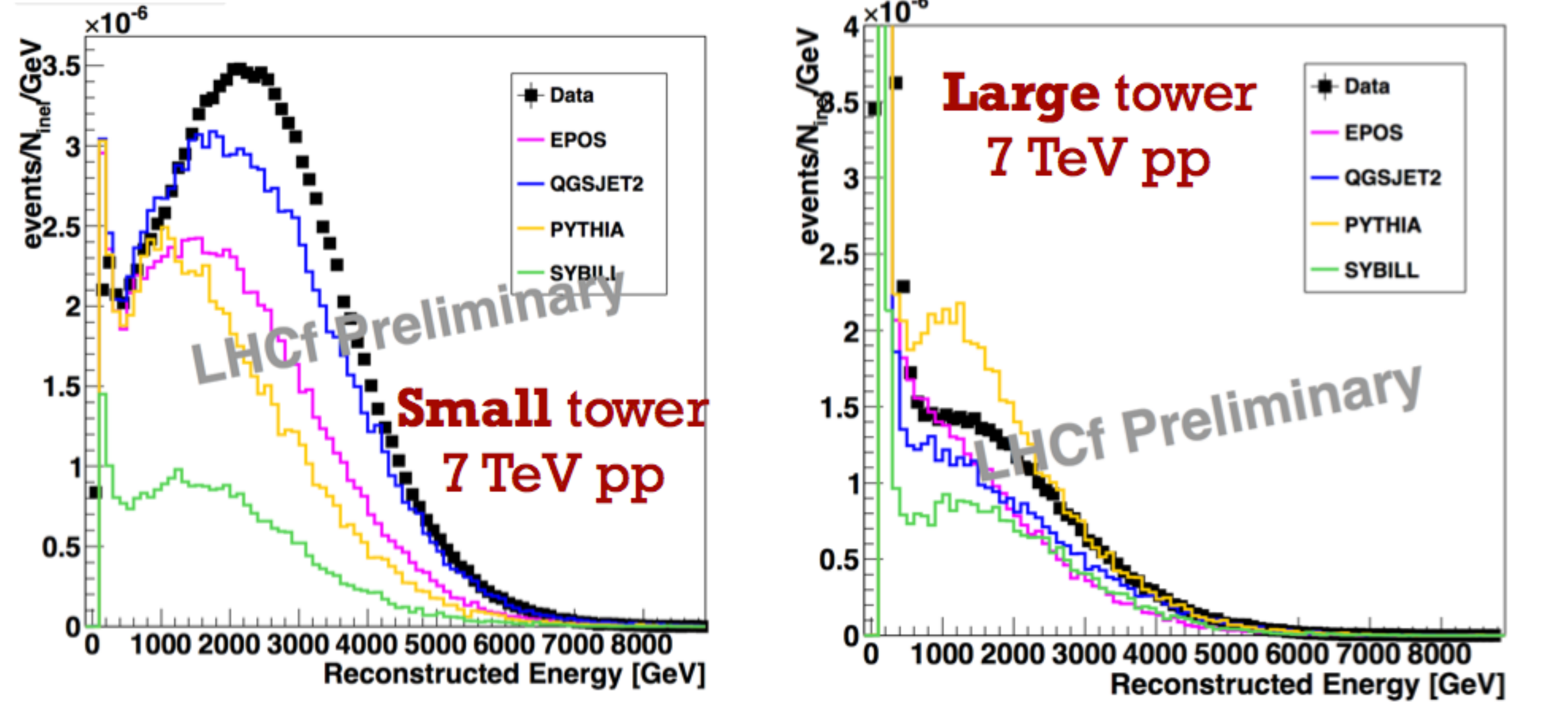}
\caption{Preliminary neutron spectra measured by LHCf in the $\sqrt{s}$ = 7 TeV p-p collisions in the small and large tower.} 
\label{fig2} 
\end{center}
\end{figure}

Figure~\ref{fig2} shows the preliminary energy spectra of forward neutrons measured by the Arm1 detector compared with the MC predictions; left and right panels correspond to the spectra measured on the small and large tower, respectively.
The effect of the limited energy resolution for hadrons ($\simeq 30 \%$) is evident from these plots. A detailed unfolding procedure, currently underway,  is hence necessary to disentangle the detector effects from the real differences in the spectra expected from the various models.

\section{Preliminary results in p-pb collisions}
\begin{figure}[htbp] 
\begin{center}
   \includegraphics[width=0.7\textwidth]{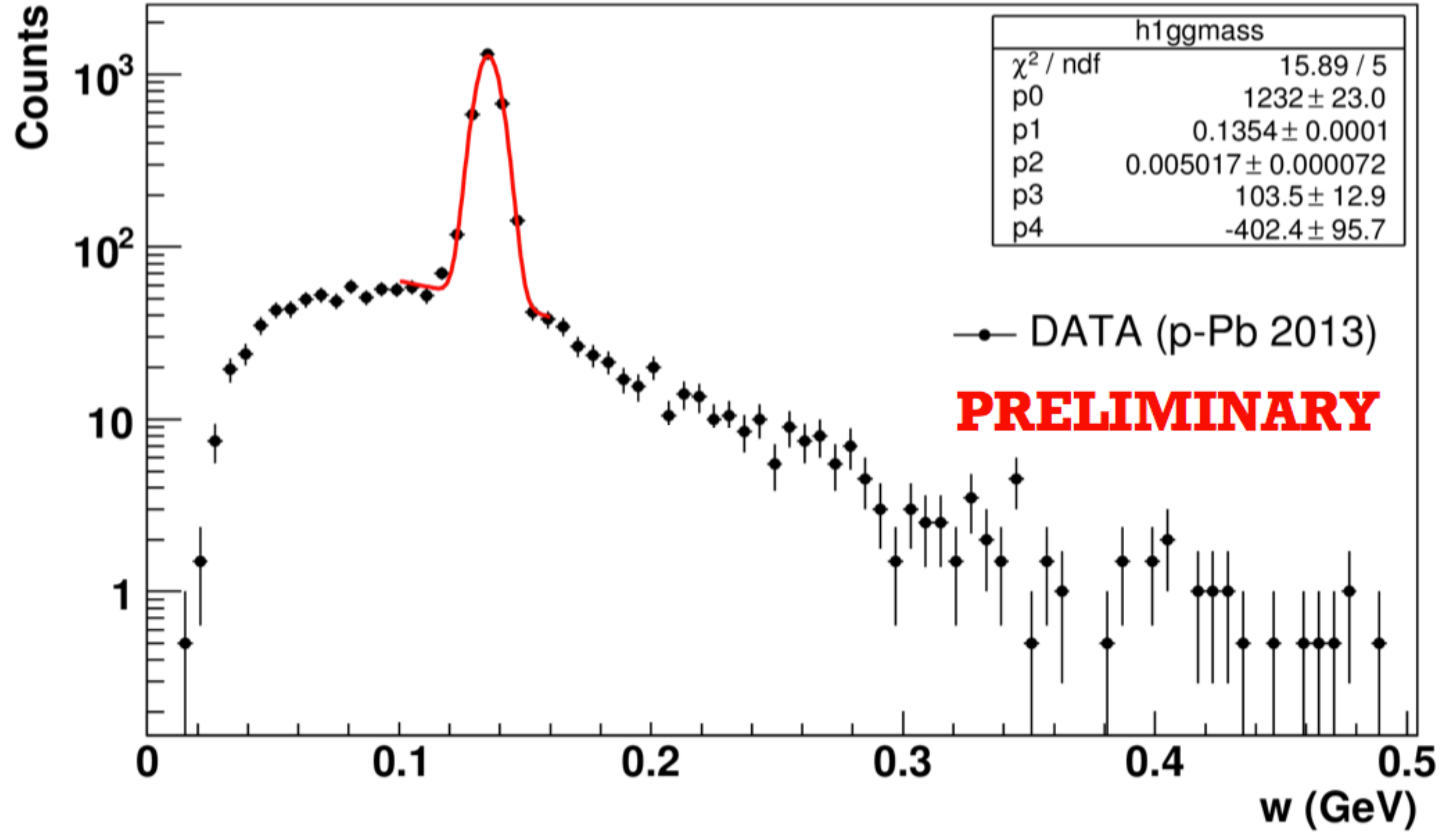}
\caption{Invariant mass spectra reconstructed in the p-Pb collisions in the proton remnant side.} 
\label{fig3} 
\end{center}
\end{figure}

At the beginning of 2013, LHCf took data in the p-Pb collisions  for a center-of-mass collision energy per nucleon of $\sqrt{s_{NN}}$ 5 TeV. Only the Arm2 detector was used in this run. In most of the operation time, the detector was located on the p-remnant side. The physics aim of the p-Pb measurement is the investigation of the nuclear effect on the production of forward energetic particles, that is one of the important effects for modeling the interactions between cosmic-rays and atmosphere. The run was very successful, and the quality of data excellent, as demonstrated by the very clean $\pi^0$ invariant mass peak shown in Figure~\ref{fig3}.

 The analysis of these data is underway, to obtain $\gamma$, $\pi^0$ and neutron transverse momentum spectra in different rapidity bins and their impact point distributions. 
 
 \vspace{3cm}
% ****************************************************************************
% BIBLIOGRAPHY AREA
% ****************************************************************************

\begin{footnotesize}
% IF YOU DO NOT USE BIBTEX, USE THE FOLLOWING SAMPLE SCHEME FOR THE REFERENCES
% ----------------------------------------------------------------------------

\end{footnotesize}
\end{document}